\documentstyle[12pt]{article}

\def\mathfrak{\bf}

\def\be{\begin{equation}}
\def\ee{\end{equation}}
\def\bea{\begin{eqnarray}}
\def\eea{\end{eqnarray}}

\def\dt#1{\on{\hbox{\bf .}}{#1}}                
\def\Dot#1{\dt{#1}}
\def\IR{\relax{\rm I\kern-.18em R}}
\def\binomial#1#2{\left(\,{\buildrel
{\raise4pt\hbox{$\displaystyle{#1}$}}\over
{\raise-6pt\hbox{$\displaystyle{#2}$}}}\,\right)}

\def\[{\lfloor{\hskip 0.35pt}\!\!\!\lceil}
\def\]{\rfloor{\hskip 0.35pt}\!\!\!\rceil}

\newcommand{\AmS}{{\protect\the\textfont2
  A\kern-.1667em\lower.5ex\hbox{M}\kern-.125emS}}

\catcode`@=11
\def\un#1{\relax\ifmmode\@@underline#1\else
        $\@@underline{\hbox{#1}}$\relax\fi}
\catcode`@=12

\def\fracm#1#2{\hbox{\large{${\frac{{#1}}{{#2}}}$}}}

\def\ad{{\kern0.5pt
                   \alpha \kern-5.05pt
\raise5.8pt\hbox{$\textstyle.$}\kern
0.5pt}}

\def\Dot#1{{\kern0.5pt
     {#1} \kern-5.05pt \raise5.8pt\hbox{$\textstyle.$}\kern
0.5pt}}

\def\a{\alpha}
\def\b{\beta}
\def\c{\chi}
\def\d{\delta}
\def\e{\epsilon}

\def\g{\gamma}
\def\h{\eta}

\def\j{\psi}

\def\l{\lambda}

\def\o{\omega}

\def\s{\sigma}

\def\z{\zeta}

\def\L{\Lambda}

\def\bo{{\raise.15ex\hbox{\large$\Box$}}}               
\def\pa{\partial}                                       
\def\TH{{\raise.2ex\hbox{$\displaystyle \bigodot$}\mskip-4.7mu \llap H
\;}}
\def\face{{\raise.2ex\hbox{$\displaystyle \bigodot$}\mskip-2.2mu \llap
{$\ddot
        \smile$}}}                                      

\def\Tilde#1{\widetilde{#1}}                    
\def\Bar#1{\overline{#1}}                       
\def\leftrightarrowfill{$\mathsurround=0pt \mathord\leftarrow \mkern-6mu
        \cleaders\hbox{$\mkern-2mu \mathord- \mkern-2mu$}\hfill
        \mkern-6mu \mathord\rightarrow$}
\def\dvec#1{\vbox{\ialign{##\crcr
        \leftrightarrowfill\crcr\noalign{\kern-1pt\nointerlineskip}
        $\hfil\displaystyle{#1}\hfil$\crcr}}}           
\def\dt#1{{\buildrel {\hbox{\LARGE .}} \over {#1}}}     

\def\fracm#1#2{\hbox{\large{${\frac{{#1}}{{#2}}}$}}}
\def\frac#1#2{{\textstyle{#1\over\vphantom2\smash{\raise.20ex
        \hbox{$\scriptstyle{#2}$}}}}}                   
\def\sfrac#1#2{{\vphantom1\smash{\lower.5ex\hbox{\small$#1$}}\over
        \vphantom1\smash{\raise.4ex\hbox{\small$#2$}}}} 
\def\bfrac#1#2{{\vphantom1\smash{\lower.5ex\hbox{$#1$}}\over
        \vphantom1\smash{\raise.3ex\hbox{$#2$}}}}       
\def\afrac#1#2{{\vphantom1\smash{\lower.5ex\hbox{$#1$}}\over#2}}    
\def\on#1#2{\mathop{\null#2}\limits^{#1}}               

\newskip\humongous \humongous=0pt plus 1000pt minus 1000pt

\newif\ifdtup

  \def\pp{{\mathchoice
          {
              \kern 1pt%
              \raise 1pt
              \vbox{\hrule width5pt height0.4pt depth0pt
                    \kern -2pt
                    \hbox{\kern 2.3pt
                          \vrule width0.4pt height6pt depth0pt
                          }
                    \kern -2pt
                    \hrule width5pt height0.4pt depth0pt}%
                    \kern 1pt
           }
            {
              \kern 1pt%
              \raise 1pt
              \vbox{\hrule width4.3pt height0.4pt depth0pt
                    \kern -1.8pt
                    \hbox{\kern 1.95pt
                          \vrule width0.4pt height5.4pt depth0pt
                          }
                    \kern -1.8pt
                    \hrule width4.3pt height0.4pt depth0pt}%
                    \kern 1pt
            }
            {
              \kern 0.5pt%
              \raise 1pt
              \vbox{\hrule width4.0pt height0.3pt depth0pt
                    \kern -1.9pt  
                    \hbox{\kern 1.85pt
                          \vrule width0.3pt height5.7pt depth0pt
                          }
                    \kern -1.9pt
                    \hrule width4.0pt height0.3pt depth0pt}%
                    \kern 0.5pt
            }
            {
              \kern 0.5pt%
              \raise 1pt
              \vbox{\hrule width3.6pt height0.3pt depth0pt
                    \kern -1.5pt
                    \hbox{\kern 1.65pt
                          \vrule width0.3pt height4.5pt depth0pt
                          }
                    \kern -1.5pt
                    \hrule width3.6pt height0.3pt depth0pt}%
                    \kern 0.5pt
            }
        }}

  \def\mm{{\mathchoice
                       {
                             \kern 1pt
               \raise 1pt    \vbox{\hrule width5pt height0.4pt depth0pt
                                  \kern 2pt
                                  \hrule width5pt height0.4pt depth0pt}
                             \kern 1pt}
                       {
                            \kern 1pt
               \raise 1pt \vbox{\hrule width4.3pt height0.4pt depth0pt
                                  \kern 1.8pt
                                  \hrule width4.3pt height0.4pt depth0pt}
                             \kern 1pt}
                       {
                            \kern 0.5pt
               \raise 1pt
                            \vbox{\hrule width4.0pt height0.3pt depth0pt
                                  \kern 1.9pt
                                  \hrule width4.0pt height0.3pt depth0pt}
                            \kern 1pt}
                       {
                           \kern 0.5pt
             \raise 1pt  \vbox{\hrule width3.6pt height0.3pt depth0pt
                                  \kern 1.5pt
                                  \hrule width3.6pt height0.3pt depth0pt}
                           \kern 0.5pt}
                       }}

\def\pd{{\kern0.5pt
                   + \kern-5.05pt \raise5.8pt\hbox{$\textstyle.$}\kern
0.5pt}}

\def\pmd{{\kern0.5pt
                  \pm \kern-5.05pt \raise6.3pt\hbox{$\textstyle.$}\kern1.5pt}}

\def\md{{\mathchoice
   {
      {{\kern 1pt - \kern-6.2pt \raise5pt\hbox{$\textstyle.$}\kern 1pt}}}
    {
      {{\kern 1pt - \kern-6.2pt \raise5pt\hbox{$\textstyle.$}\kern 1pt}}}
    {
      {\kern0.5pt - \kern-5.05pt \raise3.4pt\hbox{$\textstyle.$}\kern0.5pt}}
    {
      {\kern0.5pt - \kern-5.05pt \raise3.4pt\hbox{$\textstyle.$}\kern0.5pt}}}}

\def\ad{{\dot{\alpha}}}

\def\pp{{\mathchoice
          {
              \kern 1pt%
              \raise 1pt
              \vbox{\hrule width5pt height0.4pt depth0pt
                    \kern -2pt
                    \hbox{\kern 2.3pt
                          \vrule width0.4pt height6pt depth0pt
                          }
                    \kern -2pt
                    \hrule width5pt height0.4pt depth0pt}%
                    \kern 1pt
           }
            {
              \kern 1pt%
              \raise 1pt
              \vbox{\hrule width4.3pt height0.4pt depth0pt
                    \kern -1.8pt
                    \hbox{\kern 1.95pt
                          \vrule width0.4pt height5.4pt depth0pt
                          }
                    \kern -1.8pt
                    \hrule width4.3pt height0.4pt depth0pt}%
                    \kern 1pt
            }
            {
              \kern 0.5pt%
              \raise 1pt
              \vbox{\hrule width4.0pt height0.3pt depth0pt
                    \kern -1.9pt  
                    \hbox{\kern 1.85pt
                          \vrule width0.3pt height5.7pt depth0pt
                          }
                    \kern -1.9pt
                    \hrule width4.0pt height0.3pt depth0pt}%
                    \kern 0.5pt
            }
            {
              \kern 0.5pt%
              \raise 1pt
              \vbox{\hrule width3.6pt height0.3pt depth0pt
                    \kern -1.5pt
                    \hbox{\kern 1.65pt
                          \vrule width0.3pt height4.5pt depth0pt
                          }
                    \kern -1.5pt
                    \hrule width3.6pt height0.3pt depth0pt}%
                    \kern 0.5pt
            }
        }}

  \def\mm{{\mathchoice
                       {
                             \kern 1pt
               \raise 1pt    \vbox{\hrule width5pt height0.4pt depth0pt
                                  \kern 2pt
                                  \hrule width5pt height0.4pt depth0pt}
                             \kern 1pt}
                       {
                            \kern 1pt
               \raise 1pt \vbox{\hrule width4.3pt height0.4pt depth0pt
                                  \kern 1.8pt
                                  \hrule width4.3pt height0.4pt depth0pt}
                             \kern 1pt}
                       {
                            \kern 0.5pt
               \raise 1pt
                            \vbox{\hrule width4.0pt height0.3pt depth0pt
                                  \kern 1.9pt
                                  \hrule width4.0pt height0.3pt depth0pt}
                            \kern 1pt}
                       {
                           \kern 0.5pt
             \raise 1pt  \vbox{\hrule width3.6pt height0.3pt depth0pt
                                  \kern 1.5pt
                                  \hrule width3.6pt height0.3pt depth0pt}
                           \kern 0.5pt}
                       }}

\def\pd{{\kern0.5pt
                   + \kern-5.05pt \raise5.8pt\hbox{$\textstyle.$}\kern
0.5pt}}

\def\pmd{{\kern0.5pt
                  \pm \kern-5.05pt \raise6.3pt\hbox{$\textstyle.$}\kern1.5pt}}

\def\md{{\mathchoice
   {
      {{\kern 1pt - \kern-6.2pt \raise5pt\hbox{$\textstyle.$}\kern 1pt}}}
    {
      {{\kern 1pt - \kern-6.2pt \raise5pt\hbox{$\textstyle.$}\kern 1pt}}}
    {
      {\kern0.5pt - \kern-5.05pt \raise3.4pt\hbox{$\textstyle.$}\kern0.5pt}}
    {
      {\kern0.5pt - \kern-5.05pt \raise3.4pt\hbox{$\textstyle.$}\kern0.5pt}}}}

\def\dslash{\not{\hbox{\kern-2pt $\partial$}}}
\def\Dslash{\not{\hbox{\kern-4pt $D$}}}
\def\pslash{\not{\hbox{\kern-2.3pt $p$}}}
 \newtoks\slashfraction
 \slashfraction={.13}
 \def\slash#1{\setbox0\hbox{$ #1 $}
 \setbox0\hbox to \the\slashfraction\wd0{\hss \box0}/\box0 }

\font\ro=cmsy10                          
\def\kcr{{\hbox{\ro \char'170}}}                
\def\ktl{{\hbox{\ro \char'170}}}        
\def\ktr{{\hbox{\ro \char'170}}}        
\def\kbl{{\hbox{\ro \char'170}}}        
\def\kbr{{\hbox{\ro \char'170}}}        

\def\plpl{\raise-2pt\hbox{$\raise3pt\hbox{$_+$}\hskip-6.67pt\raise0.0pt
\hbox{$^+$}\hskip 0.01pt$}}
\def\mimi{\raise-2pt\hbox{$\raise3pt\hbox{$_-$}\hskip-6.67pt\raise0.0pt
\hbox{$^-$}\hskip 0.01pt$}}

\def\bo{{\raise.15ex\hbox{\large$\Box$}}}               
\def\pa{\partial}                                       
\def\TH{{\raise.2ex\hbox{$\displaystyle \bigodot$}\mskip-4.7mu \llap H \;}}
\def\face{{\raise.2ex\hbox{$\displaystyle \bigodot$}\mskip-2.2mu \llap {$\ddot
        \smile$}}}                                      

\def\Tilde#1{\widetilde{#1}}                    
\def\Bar#1{\overline{#1}}                       
\def\leftrightarrowfill{$\mathsurround=0pt \mathord\leftarrow \mkern-6mu
        \cleaders\hbox{$\mkern-2mu \mathord- \mkern-2mu$}\hfill
        \mkern-6mu \mathord\rightarrow$}
\def\dvec#1{\vbox{\ialign{##\crcr
        \leftrightarrowfill\crcr\noalign{\kern-1pt\nointerlineskip}
        $\hfil\displaystyle{#1}\hfil$\crcr}}}           
\def\dt#1{{\buildrel {\hbox{\LARGE .}} \over {#1}}}     

\def\fracm#1#2{\hbox{\large{${\frac{{#1}}{{#2}}}$}}}
\def\frac#1#2{{\textstyle{#1\over\vphantom2\smash{\raise.20ex
        \hbox{$\scriptstyle{#2}$}}}}}                   
\def\sfrac#1#2{{\vphantom1\smash{\lower.5ex\hbox{\small$#1$}}\over
        \vphantom1\smash{\raise.4ex\hbox{\small$#2$}}}} 
\def\bfrac#1#2{{\vphantom1\smash{\lower.5ex\hbox{$#1$}}\over
        \vphantom1\smash{\raise.3ex\hbox{$#2$}}}}       
\def\afrac#1#2{{\vphantom1\smash{\lower.5ex\hbox{$#1$}}\over#2}}    
\def\on#1#2{\mathop{\null#2}\limits^{#1}}               

\parskip=\medskipamount                 
\thicklines                         

\def\oldheadpic{                                
        \setlength{\unitlength}{.4mm}
        \thinlines
        \par
        \begin{picture}(349,16)
        \put(325,16){\line(1,0){4}}
        \put(330,16){\line(1,0){4}}
        \put(340,16){\line(1,0){4}}
        \put(335,0){\line(1,0){4}}
        \put(340,0){\line(1,0){4}}
        \put(345,0){\line(1,0){4}}
        \put(329,0){\line(0,1){16}}
        \put(330,0){\line(0,1){16}}
        \put(339,0){\line(0,1){16}}
        \put(340,0){\line(0,1){16}}
        \put(344,0){\line(0,1){16}}
        \put(345,0){\line(0,1){16}}
        \put(329,16){\oval(8,32)[bl]}
        \put(330,16){\oval(8,32)[br]}
        \put(339,0){\oval(8,32)[tl]}
        \put(345,0){\oval(8,32)[tr]}
        \end{picture}
        \par
        \thicklines
        \vskip.2in}
\def\oldtitle#1#2#3#4{\oldheadpic\begin{center}\vglue.5in{\large\bf #1}\\[.6in]
        {#2}\\[.1in] {\it Department of Physics and Astronomy}\\
        {\it University of Maryland, College Park, MD 20742}\\[.6in]
        Physics Publication \#{#3}\\ {#4}\\[1.5in] {\bf ABSTRACT}\\[.1in]
        \end{center} \begin{quotation}}                 
\def\oldTitle#1#2#3#4#5#6#7{\oldheadpic\begin{center} \vglue .4in
        {\large\bf #1}\\[.4in]
        {#2}\\[.1in] {\it Department of Physics and Astronomy}\\
        {\it University of Maryland, College Park, MD 20742}\\[.1in]
        {#3}\\[.1in] {\it {#4}}\\ {\it {#5}}\\[.4in]
        Physics Publication \#{#6}\\ {#7}\\[.5in] {\bf ABSTRACT}\\[.1in]
        \end{center} \begin{quotation}}                 
\def\border{                                            
        \setlength{\unitlength}{1mm}
        \newcount\xco
        \newcount\yco
        \xco=-21
        \yco=12
        \begin{picture}(140,0)
        \put(\xco,\yco){$\ktl$}
        \advance\yco by-1
        {\loop
        \put(\xco,\yco){$\kcr$}
        \advance\yco by-2
        \ifnum\yco>-240
        \repeat
        \put(\xco,\yco){$\kbl$}}
        \xco=158
        \yco=12
        \put(\xco,\yco){$\ktr$}
        \advance\yco by-1
        {\loop
        \put(\xco,\yco){$\kcr$}
        \advance\yco by-2
        \ifnum\yco>-240
        \repeat
        \put(\xco,\yco){$\kbr$}}
        \put(-20,13){\tiny University of Maryland Elementary Particle
Physics University of Maryland Elementary Particle Physics University of
Maryland Elementary Particle Physics}
        \put(-20,-241.5){\tiny University of Maryland Elementary
Particle Physics University of Maryland Elementary Particle Physics
University of Maryland Elementary Particle Physics}
        \end{picture}
        \par\vskip-8mm}
\def\bordero{                                           
        \setlength{\unitlength}{1mm}
        \newcount\xco
        \newcount\yco
        \xco=-31
        \yco=12
        \begin{picture}(140,0)
        \put(\xco,\yco){$\ktl$}
        \advance\yco by-1
        {\loop
        \put(\xco,\yco){$\kclr}
        \advance\yco by-2
        \ifnum\yco>-240
        \repeat
        \put(\xco,\yco){$\kbl$}}
        \xco=151
        \yco=12
        \put(\xco,\yco){$\ktr$}
        \advance\yco by-1
        {\loop
        \put(\xco,\yco){$\kcr$}
        \advance\yco by-2
        \ifnum\yco>-240
        \repeat
        \put(\xco,\yco){$\kbr$}}
        \put(-20,12){\ooo bacdefghidfghghdhededbihdgdfdfhhdheidhdhebaaahjhhdahba

hgdedge
   hgfdiehhgdigicba}
        \put(-20,-241.5){\ooo ababaighefdbfghgeahgdfgafagihdidihiidhiagfedhadbfd

ecdcdfa
   gdcbhaddhbgfchbgfdacfediacbabab}
        \end{picture}
        \par\vskip-8mm}
\def\headpic{                                           
        \indent
        \setlength{\unitlength}{.4mm}
        \thinlines
        \par
        \begin{picture}(29,16)
        \put(165,16){\line(1,0){4}}
        \put(170,16){\line(1,0){4}}
        \put(180,16){\line(1,0){4}}
        \put(175,0){\line(1,0){4}}
        \put(180,0){\line(1,0){4}}
        \put(185,0){\line(1,0){4}}
        \put(169,0){\line(0,1){16}}
        \put(170,0){\line(0,1){16}}
        \put(179,0){\line(0,1){16}}
        \put(180,0){\line(0,1){16}}
        \put(184,0){\line(0,1){16}}
        \put(185,0){\line(0,1){16}}
        \put(169,16){\oval(8,32)[bl]}
        \put(170,16){\oval(8,32)[br]}
        \put(179,0){\oval(8,32)[tl]}
        \put(185,0){\oval(8,32)[tr]}
        \end{picture}
        \par\vskip-6.5mm
        \thicklines}
\def\title#1#2#3#4{\border\headpic {\hbox to\hsize{#4 \hfill UMDEPP #3}}\par
        \begin{center} \vglue .5in {\large\bf #1}\\[.6in]
        {#2}\\[.1in] {\it Department of Physics and Astronomy}\\
        {\it University of Maryland, College Park, MD 20742}\\[1.5in]
        {\bf ABSTRACT}\\[.1in] \end{center} \begin{quotation}}  
\def\Title#1#2#3#4#5#6#7{\border\headpic
        {\hbox to\hsize{#7 \hfill UMDEPP #6}}\par
        \begin{center} \vglue .4in {\large\bf #1}\\[.4in]
        {#2}\\[.1in] {\it Department of Physics and Astronomy}\\
        {\it University of Maryland, College Park, MD 20742}\\[.1in]
        {#3}\\[.1in] {\it {#4}}\\ {\it {#5}}\\[.5in] {\bf ABSTRACT}\\[.1in]
        \end{center} \begin{quotation}}                 
\def\endtitle{\end{quotation}\newpage}                  

\def\qd{{\kern0.5pt
                   q \kern-5.05pt \raise5.8pt\hbox{$\textstyle.$}\kern
0.5pt}}

\begin{document}

\def\dt#1{\on{\hbox{\bf .}}{#1}}                
\def\Dot#1{\dt{#1}}

\def\gfrac#1#2{\frac {\scriptstyle{#1}}
        {\mbox{\raisebox{-.6ex}{$\scriptstyle{#2}$}}}}
\def\gg{{\hbox{\sc g}}}
\border\headpic {\hbox to\hsize{March 2002 \hfill
{UMDEPP 02-031}}} \par
{\hbox to\hsize{$~$ \hfill
{CALT-68-2368}}}
\par
\setlength{\oddsidemargin}{0.3in}
\setlength{\evensidemargin}{-0.3in}
\begin{center}
\vglue .10in
{\large\bf New 4D, $N$ = 1 Superfield Theory: Model of Free Massive
Superspin--${\fracm 32}$ Multiplet}\footnote{
Supported in part by National Science Foundation Grants 
PHY-01-5-23911}
\\[.5in]
I.L. Buchbinder$^a$\footnote{joseph@tspu.edu.ru},
S.\ James Gates, Jr.\footnote{On Sabbatical leave at
the California Inst. of Technology, Sept. 2001 thru July 2002}${}^,$
\footnote{gatess@wam.umd.edu}, W. D. Linch,
III$^b$\footnote{linch@bouchet.physics.umd.edu} and J.
Phillips$^b$\footnote{ferrigno@physics.umd.edu}
\\[0.06in]
{\it {}$^a$Department of Theoretical Physics, Tomsk State Pedagogical
University\\ 634041 Tomsk, Russia}
\\[0.05in]
{\it {}$^b$Department of Physics, University of Maryland,
College Park\\
MD 20742-4111 USA}
\\[1.0in]

{\bf ABSTRACT}\\[.01in]
\end{center}
\begin{quotation}
{ We present a Lagrangian formulation for the free superspin-${\fracm 32}$
massive 4D, $N$=1 superfield. The model is described by a dynamical real 
vector superfield and an auxiliary real scalar superfield. The 
corresponding multiplet contains spin-1, spin-2 and two
spin-${\fracm 32}$ propagating component fields on-shell. The auxiliary
superfield is needed to ensure that the on-shell vector superfield
carries only the irreducible representation of the Poincar\'e supergroup
with a given mass and the fixed superspin of ${\fracm 32}$. The 
bosonic sector of the component level of the model is also presented.}

${~~~}$ \newline

\endtitle

\noindent
\noindent{\bf 1.} The problem of constructing consistent Lagrangian
formulations for higher spin fields has attracted significant attention
and interest for a long time.  Formulations for {\it {non-supersymmetric}}
free massive and massless higher spin theories have been investigated
thoroughly and are well understood. Supersymmetric extensions of massless
higher spin models have been developed.  However, the problem of
constructing supersymmetric {\it {massive}} higher spin actions in
Minkowski space is still open. In this letter we begin a study of such
actions and present a theory that can be viewed as a supersymmetric
extension of the free massive spin- 2 model given by Fierz and Pauli [1].

A general solution for {\it {non}}-{\it {supersymmetric}} Lagrangians of
the free higher spin massless [2] and massive [3] field theories in
Minkowski space was given over 30 years after the pioneering work of
Fierz and Pauli.  Further development of massless theories was related to
free theories in Minkowski [4] and constant curvature spaces [5]. 
Construction of interacting theories was discussed in [6] (see also the
reviews [7]).  Progress for massive theories was restricted mainly to the
spin-2 field in AdS space [8-11], and only recently Lagrangians for
arbitrary integer spin fields in constant curvature spaces were found
[12].

In the supersymmetric case, the problem of writing equations of motion 
for massless higher spin supermultiplets in Minkowski 4D, $N$ = 1
superspace was first described using field strength superfields [13]. 
Later the problem of writing superspace Lagrangians, consistent with
the field strength superfield equations, for massless higher spin 
fields in Minkowski space was solved using superfield methods in 
[13,14] (see also [15]).  

Furthermore, supersymmetric massless models in AdS superspace have been
constructed in [16] and further investigated in [17].  These last two
investigations may prove to be precursors to an interesting recent
development within the confines of AdS/CFT duality.  It has recently
been suggested by Witten [18] that the models in the work of [6,7] can 
be interpreted as holographic duals of $g^2 N$ = 0 supersymmetric 4D, 
$N$ = 4 SU$(N)$ Yang-Mills theory as $N \to \infty$.  As noted in
[17], the spectrum of fields found in this superspace approach coincides
with the components required in the Fradkin-Vasiliev approach [6,7].

As surprising as it may seem and as far as we can tell, a general massive supersymmetric irreducible higher spin action for 4D, $N=1$ theory has never been presented in the literature. The only references of which we are aware \cite{berkovits} are devoted to the derivation of the action for the first massive state from open superstring field theory (which contains the massive superspin-${\fracm 32}$ multiplet in addition to two massive scalar multiplets and hence describes a reducible representation of the Poincare supergroup). \footnote{We are very grateful to N. Berkovits for directing our attention to the papers \cite{berkovits}.} In this paper we carry out the detailed study of the superfield model corresponding to the most general $N=1$ supersymmetric extension of the Fierz-Pauli theory.
We regard this as simply the first step in a long term program to
extend the generating function techniques of [17] to cover the
arbitrary spin massive supermultiplets in Minkowski space.  A most
important goal of such studies is to understand the possibility of
uncovering a Higgs-like mechanism that connects the massless theories
[13,14,15] to the massive ones that are the logical extension of
our present investigation.

It is easy to understand why constructing massive supersymmetric higher
spin theories at the level of actions is more difficult than constructing
massless ones.  Massless theories possess gauge invariances which impose
rigid restrictions on the form of the Lagrangian.  This is why the form
of the Lagrangian for massive arbitrary spin fields found in [3] looks
much more complicated than that of the massless case [2]. Another
important difference between massive and massless higher spin theories is
the number of auxiliary fields which are needed in order for the
equations of motion to be compatible with irreducible representations of
the Poincar\'e group.  The necessity of such auxiliary fields was first
pointed out by Fierz and Pauli [1].  For example, in the case of a
massive integer spin-$s$ theory  we have to consider auxiliary fields
with all spins beginning with $s-2$ and below.  In the massless case, 
the description of an integer spin-$s$ field demands only one auxiliary
field with spin $s-2$. Therefore, it is natural to expect that a
supersymmetric massive higher spin theory will contain such auxiliary
fields, and that these fields must form supermultiplets in order to
preserve the supersymmetry.

At first sight, the construction of supersymmetric free massive
higher spin theories seems to be a trivial task. It is sufficient
to take a suitable number of bosonic and fermionic Lagrangians given in
[3]. However, by proceeding in this manner we can only expect to obtain
a theory with on-shell supersymmetry.  In such a theory, neither the
auxiliary\footnote {A feature of massive higher spin theories is the
presence of specific auxiliary fields ensuring \newline ${~~~~\,}$
a Lagrangian formulation. In the supersymmetric case, we have to expect 
an appearance of\newline ${~~~~\,}$ the corresponding auxiliary
superfields. Of course, these auxiliary superfields, allowing a La-
\newline ${~~~~\,}$ grangian formulation, are ``auxiliary'' in a
completely different  sense in comparison with the \newline ${~~~~\,}$
auxiliary component fields responsible for realizing off-shell
supersymmetry} fields responsible for off-shell supersymmetry nor the
supersymmetry transformations are known. Searching for these fields and
discovering the supersymmetry tranformations becomes the main problem. 
Clearly the most adequate way to develop off-shell supersymmetric
theories must be based on superfield methods.  In the superfield
formalism, off-shell supersymmetry and supersymmetry transformations are
built in, and the problem is only to construct the superfield Lagrangian.

To construct a supersymmetric Lagrangian formulation for
massive higher spin fields we begin with superfields carrying
the irreducible massive representations of the Poincar\'e supergroup.
Such representations are described by the mass $m$ and superspin $Y$
[15].  The irreducible representation  with given $m$ and $Y$ is a
multiplet containing irreducible representations of the conventional
Poincar\'e group with the same mass $m$ and with ordinary spins
$(Y-1/2,~Y,~Y,~Y+1/2)$ [15]. As a result, if we construct a Lagrangian
formulation for a superfield with superspin $Y$, we automatically
obtain the  Lagrangian formulation for the above multiplet of the
conventional fields together with all proper auxiliary fields
responsible for off-shell supersymmetry.

In this letter we construct a superfield Lagrangian corresponding to,
perhaps, the simplest 4D, $N$ = 1 supersymmetric massive higher spin
model, a free theory for the superspin-${\fracm 32}$ superfield.   The
corresponding on-shell supermultiplet contains four particles: one
particle with spin-2, two particles with spin-${\fracm 32}$ and one
particle with spin-1. This theory can be viewed as a supersymmetric
generalization of the Fierz-Pauli model.

\noindent{ \bf 2.} A convenient theory of projection operators required
for obtaining irreducible 4D, $N$ = 1 supermultiplets was developed long
ago [19] and subsequently modified (see the details in [15]) to describe
the massive case.  We begin with a brief description of the massive
irreducible superspin-${3\over 2}$ representation of the Poincar\'e
supergroup. Such a representation is realized in a linear space of the
real vector $N$=1 superfield $V_{\un a}$. This is the simplest superfield
containing a spin-2 component field. To obtain the massive spin-2 
representation this superfield $V_{\un a}$ must satisfy the on-shell 
conditions\footnote{In this equation and throughout this presentation, 
we use {\it{Superspace}} [13] conventions.  In \newline ${~~~~\,}$ particular, 
the underline vector index simultaneously denotes the usual Minkowski 
4-vector \newline ${~~~~\,}$ index as well as a pair of undotted-dotted Weyl spinor
indices.  See ref. [13] for details.}
\bea
\label{massshell}
(\Box\,-\,m^2)\, V_{\un a} ~=~0~~~,~~~\partial^{\un a} \, V_{\un a}~=~0
~~~.
\eea
We also
require that $V_{\un a}$ forms an irreducible representation in the space
of superfields.  To satisfy this requirement $V_{\un a}$ must be either
chiral or linear.  To be sure that we have a superspin-${\fracm 32}$
superfield, we will need the superspin operator ${\bf C}$.  As
described in [15], ${\bf C}$ takes the form
\bea
{\bf C}\, V_{\un a} ~=~ m^4\{~ 2I \,+\, ({\fracm 34} \,+\, {\bf 
B}){\cal P}_{(0)} ~\} \, V_{\un a} ~~~,
\eea
where the linear subspace projectors ${\cal P}_{(0)}$ and ${\bf B}$ are
given by:
\bea
{\cal P}_{(0)}&=&-{1\over 8m^2}D^{\a}{\Bar D}{}^2 D_{\a} ~~~, \cr 
{\bf B}&=&{1\over{4m^2}}({\cal M}{}_{\a\b}{\bf P}^{\b}_{~\Dot\a}
\,-\, {\Bar
{\cal M}}{}_{\Dot\a\Dot\b}{\bf P}^{~\Dot\b}_{\a})[D^{\a},{\Bar
D}{}^{\Dot\a}] ~~~.
\eea
By acting on $V_{\un a}$ with the superspin operator we can determine the
differential constraints necessary to produce a superspin-${\fracm 32}$
representation.  A superspin-1 representation would be obtained if $V_{\un a}$ 
were chiral, since ${\cal P}_{(0)}V_{\un a}=0$ and ${\bf C}V_{\un a}=1(1+1)m^4
V_{\un a}$.  If $V_{\un a}$ is a linear superfield then ${\cal P}_{(0)}
V_{\un a}=V_{\un a}$ and:
\bea
{\bf B}V_{\un c} ~=~ {i\over2m^2}\, [~ \partial_{\g\Dot\a}{\Bar
D}{}^{\Dot\a}D^{\a}V_{\a\Dot\g} \,+\, \partial_{\Dot\g\a}D^{\a}{\Bar
D}{}^{\Dot\a}V_{\g\Dot\a} \,-\, 2im^2V_{\un c} ~]  ~~~,
\eea
If we set:
\bea
\label{irred}
D^{\a}V_{\un a} ~=~ 0~~~,~~~ {\Bar D}{}^{\Dot\a}V_{\un a}~=~0 ~~~,
\eea
then ${\bf B}V_{\un a}=V_{\un a}$, and ${\bf C}V_{\un a}=m^4{{15}\over 4}
V_{\un a} = m^4{3\over2}({3\over 2}+1)V_{\un a}$.  Thus, we
have a superspin-${3\over 2}$ irreducible representation.  As a result 
we have shown, if $V_{\un a}$ satisfies (\ref{massshell}) and (\ref{irred}) on-shell, it carries
out a massive superspin-${3\over 2}$ irreducible representation of the
Poincar\'e supergroup.


\noindent{\bf 3.} Our purpose is to derive a superfield Lagrangian
leading to such equations of motion for $V_{\un a}$ which reproduce 
on-shell conditions (\ref{massshell}) and (\ref{irred}). We show that this is not possible, 
if $V_{\un a}$ is the only variable in the Lagrangian.  We proceed to 
introduce the auxiliary scalar superfield $V$ to construct a 
Lagrangian with the above properties.

To begin, we consider the superfield $V_{\un a}$ as an arbitrary real
superfield having no special properties.  We demand that the Lagrangian
formulation reproduce conditions (\ref{massshell}) and (\ref{irred}) solely as a consequence of
the equations of motion.  In this way, $V_{\un a}$ will form a massive
superspin-${3\over 2}$ irreducible representation of the Poincar\'e
supergroup on-shell.  With this in mind, we give the most general
quadratic form of the superfield Lagrangian constructed from $V_{\un 
a}$ and covariant derivatives.  The Lagrangian contains some arbitrary 
coefficients which are fixed by requiring that the equations of motion 
reproduce the conditions (\ref{massshell}) and (\ref{irred}).  The following identity and its 
complex conjugate will be used further
\bea
\partial_{\b\Dot\a}V^{\un b} ~=~ -\partial^{\un b}V_{\b\Dot\a} ~+~
 (\pa^{\un c} V_{\un c}) \d_{\Dot\a}{}^{\Dot \b} ~~~.
\eea

The most general quadratic action constructed from $V_{\un a}$ and
supercovariant derivatives has the form
\bea
\label{genactVa}
{~}S_1[V_{\un a}] &=&\int d^8z ~ \Big\{ ~ 
\fracm 12 \a_1 \, V^{\un a}D^{\b}{\Bar D}{}^2D_{\b} V_{\un a} ~+~ 
\fracm 12 \a_2 \, V^{\un a}\, \Box \, V_{\un a}+~ \fracm 12 \a_3 
\,V^{\un a} \, \partial_{\un a} \, \partial^{\un b} \,  V_{\un b} 
 {~~}\cr
&~&{~~~~~~~~~~~~}~+~ \fracm 12 \a_4 \, V^{\un a} [D_{\a},{\Bar
D}{}_{\Dot\a}] [D_{\b}, {\Bar D}{}_{\Dot\b}] V^{\un b}~+~\a_5 V^{\un a}[D_\a,{\Bar D}_{\dot\a}] \pa_{\un b}V^{\un b}\cr
&~&{~~~~~~~~~~~~}~+~ \fracm 12 m
V^{\un a}(\b D^2+\b^{\ast}{\Bar D}{}^2) V_{\un a} 
\,-\, \fracm 12 m^2 V^{\un a} V_{\un a} ~ \Big\} ~~~,
\eea
where $\a_i$ and $\b$ are arbitrary dimensionless coefficients
and $m$ denotes the mass parameter. Any other term quadratic in $V_{\un a}$ can be shown to be some combination of the terms included in (\ref{genactVa}). We further note that, although $\b$ is complex, the $\a_i$ coefficients are all real.    

The equation of motion corresponding to the action (\ref{genactVa}) is
\bea
\label{EomVa1}
{{\d S_1} \over {\d V_{\un a}} } ~=~  0  &\Rightarrow&
\a_1D^{\b}{\Bar D}{}^2D_{\b} V_{\un a} \,+\, \a_2\Box
V_{\un a} \,+\, \a_3\partial_{\un a}\partial^{\un b}
V_{\un b} 
+\,  \a_4[D_{\a},{\Bar D}{}_{\Dot\a}][ D_{\b},{\Bar
D}{}_{\Dot\b}] V^{\un b} \cr
&~&+\,\a_5 \left( [D_\a,{\Bar D}_{\dot\a}]\pa_{\un b}V^{\un b}-[D_\b,{\Bar D}_{\dot\b}]\pa_{\un a} V^{\un b}\right)\cr
&~&+m(\b D^2+\b^{\ast}{\Bar D}{}^2) 
V_{\un a}- m^2 V_{\un a} = 0 \eea
The procedure for fixing the coefficients is as follows. First for convenience let us define $E_{\un a} \equiv $ 
${{\d S_1} / {\d V^{\un a}} } $ and by acting with the various
first order differential operators, ${\cal O}(\partial ,D, {\bar D}
)$, on $E_{\un a}$, we can set the coefficients so that ${\cal O}
E_{\un a}=-m^2{\cal O}V_{\un a}=0$.  Since condition (\ref{irred}) implies that
$\partial^{\un a} V_{\un a}=0$ it is easier first to attempt to set $D^{\a} V_{
\un a}=0$.  In this case, ${\cal O}^{\a}E_{\un a}=D^{\a} E_{\un
a}=-m^2D^{\a}V_{\un a}=0$ leads to $\b=0$ and the following:  
\bea
2i\a_1+6i\a_4+\a_5=0~~,~~\a_2=0~~,~~8\a_4-\a_3=0~~,~~2i\a_4-\a_5=0
\eea
Recalling the reality of these coefficients, we find that $E_{\un a}=-m^2 
V_{\un a}=0$, and there is no way to satisfy (\ref{massshell}).  Next we try to first set
$\partial^{\un a} V_{\un a}=0$ and then
$D^{\a} V_{\un a}=0$.  Here ${\cal O}E_{\un a}$ is simply the
divergence of the equation of motion which yields $\b=0$ and:
\bea
\a_1~=~ -\a_4 ~~,~~
\a_2 \,-\, 2\a_3 \,+\, 8\a_4~=~ 0 ~~,~~\a_5=0~~~.
\eea
Setting  $D^{\a}V_{\un a}=0$ leads to $\a_2=0$ and:
\bea
\a_1 ~=~ -3\a_4 ~~~,
\eea
thus $E_{\un a}=-m^2V_{\un a}=0$ again.

We see that the conditions (\ref{massshell}) and (\ref{irred}) are too constraining to be 
produced by an action quadratic in $V_{\un a}$ only.  We are forced to
introduce an auxiliary field to set $\partial^{\un a} V_{\un a}=0$.  Since this
term is a real scalar superfield it is natural to introduce a real
scalar superfield $V$ to cancel it.  Now, we seek a Lagrangian such
that the equations of motion imply $V=0$.  If this occurs, the equation
of motion of $V$ will produce a differential constraint on $V_{\un a}$.  
The most general action with $V \,V_{\un a}$ coupling terms and kinetic 
$V$ terms is:
\bea
\label{genactV}
&\,&
S_2[V_{\un a}, V] \,=\, \int d^8z ~\Big\{ 
~\g m \, V\partial^{\un a}V_{\un a} \,+\, \Tilde\g mV[D^{\a}, {\Bar D}
{}^{\Dot\a}]V_{\un a} \,+\, \fracm 12 \d_1 \, V\Box V
{~~~~~} \cr 
&~&{~~~~~~~~~~~~~~~~~~~~~~~~~~~~~~~~~~~~~~~~~~~} +\,
 \fracm 12 \d_2 \, V\{ D^2,
{\Bar D}{}^2 \}V \,+\,  \fracm 12 \d_3 \, m^2 \, V^2 ~ \Big\} ~~~.
\eea
The equation of motion for $V_{\un a}$ becomes $E_{\un a}-\g
m\partial_{\un a}V=0$ where $E_{\un a}$ is given by (\ref{EomVa1}), and for $V$:
\bea
\label{EomV1}
\g m \, \pa^{\un a} V_{\un a} \,+\, \d_1\Box V \,+\, \d_2\{ D^2, {\Bar 
D}{}^2\}V \,+\, \d_3 m^2V ~=~0 ~~~.
\eea
Here we have set $\Tilde \g=0$.  This ensures that the differential
constraint on $V_{\un a}$ implied by (\ref{EomV1}) when $V=0$ is only
$\partial^{\un a} V_{\un a}=0$.  By taking the divergence of the equation of motion
for $V_{\un a}$ (\ref{EomVa1}), and using (\ref{EomV1}) to substitute for $\partial^{\un a} V_{\un a}$, we
can force $m^2{\fracm {\d_3}{\g}}V=0$.  This requires that $\b$ and $\a_5$ vanish
once again.  We also have the following constraints:
\bea
\label{constraints}
&~&{\fracm
43}\a\d_2 ~=~ (\a_1+\a_4)\d_1 ~=~ -{\fracm 1{12}}\a\d_1 ~~~, \cr
&~&2\d_2 ~=~ (\a_1+\a_4)\d_3 ~~,~~ \d_1 \,+\, 16\d_2 \,+\, {\fracm
23}\a\d_3 ~=~ -2\g^2 ~~~,
\eea
where $\a\equiv-{\fracm 32}\a_2+3\a_3-12\a_4$.  At this point $V=0$, 
thus (\ref{EomV1}) implies $\partial^{\un a} V_{\un a}=0$.  Now we want $D^{\a}V_{\un a}
=0$ which requires:  $\a_2=0$ and $\a_1=-3\a_4$.  The equation of motion
for $V_{\un a}$ becomes, $-8\a_1\Box V_{\un a}-m^2 V_{\un a}=0$.  To get
the correct mass shell condition in (\ref{massshell}), we must have:  $\a_1=-{\fracm
18}$. All of these conditions can be solved in terms of $\a$ and $\g$
\bea
\a_3 ~=~ {\fracm 13}(\a+{\fracm 12})&,&
\d_1~=~ -2\a\g^2 ~~~, \cr
\d_2~=~{\fracm 14}(\a-{\fracm 12})\g^2 &,&
\d_3 ~=~ 3(1-2\a)\g^2 ~~~,
\eea
where $\a$ is either $0$ or $1$, which is implied by the first relation
in equation (\ref{constraints}) since $\a_1+\a_4=-{\fracm 1{12}}$ as found above. As a result all
coefficients are found and we can write down a final action as a sum 
of $S_1[V_{\un a}]$ (\ref{genactVa}) and $S_2[V_{\un a}, V]$ (\ref{genactV}) with the above
coefficients. This final action is:
\bea
\label{finact}
S[V_{\un a}, V] &=& \int d^8z 
~\Big\{ -{\fracm 1{16}} V^{\un a} D^{\b}{\Bar D}{}^2 D_{\b}
V_{\un a} \,+\, {\fracm 1{48}} V^{\un a}[D_{\a},{\Bar D}{}_{\Dot\a}]
[D_{\b},{\Bar D}{}_{\Dot\b}] V^{\un b} {~~} \cr
&~&{~~~~~~~~~~~}+\, 
{\fracm 16}(\a+{\fracm 12})
V^{\un a}\partial_{\un a}\partial^{\un b} V_{\un b}
-{\fracm {m^2}{2}} V^{\un a} V_{\un a} +
mV\partial^{\un a} V_{\un a}\,-\, \a V\Box V
\cr
&~&{~~~~~~~~~~~}+{\fracm 18}(\a-{\fracm 12})V\{ D^2, {\Bar D}{}^2 \}V+{\fracm 32}(1-2\a)m^2VV\Big\} ~~~.
\eea
It should be noted that when $\a=1$, action (\ref{finact}) reproduces the exact
kinetic terms from linearized minimal $N$=1 supergravity in the gauge
where the chiral compensator is equal to unity (see e.g. [15]). 

Thus the Lagrangian (\ref{finact}) leads to the equations of motion defining
the massive superspin-${\fracm 32}$ irreducible representation of the
Poincar\'e supergroup.  We have shown that a real scalar auxiliary
superfield ensures the existence of a Lagrangian formulation.

It is interesting to compare the action (\ref{finact}) with a superfield action derived within open superstring field theory to describe the free dynamics of the fields corresponding to the first massive open superstring level \cite{berkovits}. The action constructed in the last paper of ref. \cite{berkovits} contains the real vector superfield $V_{\un a}$ (as in the action (\ref{finact}) ) plus two real scalar superfields and a complex spinor superfield coupled to the field $V_{\un a}$ and among themselves. The equations of motion from this action lead not only to eqs. (\ref{massshell},\ref{irred}) defining a massive superspin-${\fracm 32}$ irreducible representation of the Poincar\'e supergroup but also to the equations for two massive chiral superfields. This is not a surprise since the field content of the first massive level of the open superstring corresponds to a reducible representation of the Poincar\'e supergroup and includes the superspin-${\fracm 32}$ field together with two superspin-${\fracm 12}$ fields .  The action (\ref{finact}) and the action given in \cite{berkovits} were constructed to realize two different aims, namely describing the dynamics of different numbers of degrees of freedom and therefore they have essentially different structures. One can assume that there must exist some (rather nontrivial) way to make the field redefinitions in the action given in \cite{berkovits} and decouple the contribution of the pure superspin-${\fracm 32}$ multiplet (action (\ref{finact})) from that of the two superspin-${\fracm 12}$ multiplets.

\noindent {\bf 4.} Let us analyze the massless limit of the action (\ref{finact}). To do this let us rewrite our lagrangian with $m$ set to zero as
\bea
\label{m=0act}
{\cal L}_{m\to 0}&=&-{1\over16}V^{\un a}D^\b {\bar D}^2 D_\b V_{\un a} + {1\over {48}}V^{\un a}\left[ D_\a,{\bar D}_{\dot \a}\right]\left[D_\b,{\bar D}_{\dot \b}\right] V^{\un b} \cr\cr
&~&~ +\left({1\over 4}\a+{1\over12}(1-\a)\right) V^{\un a}\pa_{\un a}\pa_{\un b}V^{\un b}+{1\over 8}\a VD^\a{\bar D}^2D_\a V \cr\cr
&~&~  -{1\over8}(1-\a)D^2V\bar D^2 V~.
\eea
As pointed out just after eq. (\ref{finact}), the choice $\a=1$ corresponds to linearized old minimal supergravity in the gauge $\s=0$ together with a decoupled U(1) theory coming from the auxiliary field $V$. The case of $\a=0$ demands a more careful treatment. Here the remnant action of the auxiliary superfield $V$ is reduced to the action of a chiral $\bar D^2V$ and anti-chiral $D^2V$ superfield. We will now show that the $V_{\un a}$ sector of both the $\a=0$ and $\a=1$ theory describe the same massless representation of supersymmetry. 

Let us consider the quantity $i\pa_{(\a}{}^{\dot \a}E_{\b)\dot\a}$, where $E_{\un a}$ is the equation of motion for the field $V_{\un a}$, now in the massless case. It is given by
\bea
i\pa_{(\a}{}^{\dot \a}E_{\b)\dot\a}&=&{i\over8}D^\g\bar D^2 \pa_{(\a}{}^{\dot \g}D_\b V_{\g)\dot\g}\equiv D^\g {\bf W}_{\a\b\g}~,
\eea
where ${\bf W}_{\a\b\g}$ is the superhelicity-${\fracm 32}$ field strength of linearized old minimal supergravity. Notice, in particular, that this quantity is independent of $\a$. Hence both theories are describing the same massless irrep of the super-Poincar\'e group. It follows from this that they are dual descriptions. This duality can be made explicit off-shell using the following action functional
\bea
\label{dualact}
A[V_{\un a},\s,\bar \s,{\bf U}]&:=&\int d^8z ~\Big\{ -{1\over16}V^{\un a}D^\b {\bar D}^2 D_\b V_{\un a} + {1\over {48}}V^{\un a}\left[ D_\a,{\bar D}_{\dot \a}\right]\left[D_\b,{\bar D}_{\dot \b}\right] V^{\un b} \cr \cr
&~&~+{1\over12} V^{\un a}\pa_{\un a}\pa_{\un b}V^{\un b} +{\bf U} \left[ \pa_{\un a}V^{\un a}+3i \left(\s-\bar \s\right)\right]+{3\over2}{\bf U}^2 \Big\}~,
\eea
which is invariant under the following gauge transformations
\bea
\d V_{\un a}&=&\bar D_{\dot \a}L_\a -D_\a \bar L_{\dot \a} \cr\cr
\d \s&=&-{1\over 12}\bar D^2 D^\a L_\a \cr\cr
\d {\bf U}&=&{i\over12}\left(D^\a\bar D^2L_\a-\bar D_{\dot \a}D^2 \bar L^{\dot \a}\right)~.
\eea
In these expressions $\s$ is a chiral superfield and ${\bf U}$ is an unconstrained real superfield. Upon varying this action with respect to ${\bf U}$ and substituting ${\bf U}$ from the resulting equation of motion, we reproduce the action for old minimal supergravity. Varying, instead, with respect to $\s$ and $\bar \s$ yields the on-shell constraints $\bar D^2 {\bf U}=0=D^2 {\bf U}$, i.e. ${\bf U}$ is a real linear superfield. Substituting this constraint cancels the $\s$ and $\bar \s$ fields resulting in the following form for the action
\bea
S[V_{\un a},{\bf U}]&:=&\int d^8z ~\Big\{ -{1\over16}V^{\un a}D^\b {\bar D}^2 D_\b V_{\un a} + {1\over {48}}V^{\un a}\left[ D_\a,{\bar D}_{\dot \a}\right]\left[D_\b,{\bar D}_{\dot \b}\right] V^{\un b} \cr \cr
&~&~~~~~~~~~~~~~~~~~~~~~~~~~  +{1\over12} V^{\un a}\pa_{\un a}\pa_{\un b}V^{\un b} +{\bf U} \pa_{\un a}V^{\un a} +{3\over2}{\bf U}^2 \Big\}~.
\eea
If we choose the gauge ${\bf U}=0$, we reproduce the $m=0$, $\a=0$ action (\ref{m=0act}) in the $V_{\un a}$ sector. One may note that the construction of the action functional (\ref{dualact}) is very similar to the construction of the action functional ensuring the duality between old minimal and new minimal linearized supergravity (see eq. (6.7.37) in \cite{Buch2}). 

This theory is dual to linearized old minimal supergravity in the same way that new minimal supergravity is dual to old minimal supergravity. In fact, we can describe this similarity more precisely as follows. The residual gauge transformations of the gauge chosen above constrain the gauge parameter superfield by 
\bea
D^\a\bar D^2 L_\a-\bar D_{\dot \a}D^2\bar L^{\dot \a}=0 ~.
\eea
This equation has the solution 
\bea
L_\a=-D_\a K-\bar D^{\dot \a}\z_{\un a}~,
\eea
in terms of which we can rewrite
\bea
\d V_{\un a}=\left[ D_\a , \bar D_{\dot \a}\right]K+\L_{\un a}+\bar \L_{\un a} ~,
\eea
with $K=\bar K$ and $\bar D_{\dot \a}\L_{\un b}=0$. This is to be compared with the case of new minimal supergravity \cite{Buch2}. We see that the solution presented here is the ``imaginary part" of what is written in new minimal supergravity in the following sense: Here $\d {\bf U}={i\over 12} {\rm Im}\left(D^\a\bar D^2L_\a\right)$ while in new minimal supergravity we have $\d{\bf U}={1\over4}{\rm Re}\left(D^\a\bar D^2L_\a\right)$. These observations suggest that there is yet another theory of (non-linear) minimal supergravity.


\noindent {\bf 5.} The superfield action (\ref{finact}) is a complete solution to
the problem under consideration. However it would be useful and
interesting to rewrite this action in terms of components of the
superfields $V_{\un a}$ and $V$ and obtain a component Lagrangian. 
We now present an analyis of the bosonic sector of this Lagrangian.

We use the
following definitions for the bosonic component fields:
\bea
\label{components}
V^{\un a}| \,=\, A^{\un a} ~~~,~~~ D^2V^{\un b}|&=&-4 
F^{\un b}~~~,~~~ {\Bar D}{}^2 V^{\un b}| \,=\, -4 {\Bar 
F}{}^{\un b} ~~~, \cr
\{ D^2,{\Bar D}{}^2 \} V^{\un a}| \,=\, 32 D^{\un a} &,& 
[ D_{\a}, {\Bar D}{}_{\Dot \a}] V^{\un b}| \,=\, 2 V_{\un 
a}^{~~\un b} ~~~, \cr 
V |=\c~~~,~~~{\rm D^2}V |&=&-4 \j ~~~,~~~ {\Bar D}{}^2V | \,=\,
-4 {\Bar \j} 
\cr
\{ D^2,{\Bar D}{}^2 \}V|=32
\h &,& [ D_{\a}, {\Bar D}{}_{\Dot \a}]V|=2 \l_{\un a} ~~~, \cr 
V^{\un a \un b}= h^{\un a \un b}&+& \o^{\un a \un b} \,+\, \fracm
14 \, \h^{\un a\un b} \, h ~~~, \cr 
h^{\un a \un b}= h^{\un b \un a}~~~,~~~ h_{\un a}^{~\un a}&=&0 ~~~,~~~
{\o^{\un a \un b}}=-{\o^{\un b \un a}} ~~~.
\eea
After integrating the action in (\ref{finact}) over the superspace Grassmann
coordinates and using (\ref{components}) this action can be rewritten in component 
form for arbitrary $\a$.
\bea
\label{compact}
S[V_{\un a},V] &=& \int d^4x ~\Big\{
-{\fracm 1{24}}(\a+2)\partial_{\un b}    h^{\un b \un a}\partial^{\un c} h_{\un
c \un a} 
\,+\, {\fracm 18}h^{\un a \un b}\Box h_{\un a  \un b}
\,-\, {\fracm 18} m^2h^{\un a \un b} h_{\un a  \un b}\cr
&\,& {~~~~~~~~~~} -{\fracm 1{48}}\a h\Box h
\,-\, {\fracm 18}m^2h^2 \,-\,{\fracm 16}A^{\un a}   \Box^2A_{\un a}   
\,+\,{\fracm 1{12}}\a(\partial\cdot A)\Box (\partial\cdot A)\cr
&\,&  {~~~~~~~~~~} +{\fracm 14} m^2A^{\un a}\Box A_{\un a}
\,+\,{\fracm 1{24}}\o^{\un a \un b}\Box\o_{\un a \un b} \,-\, 
{\fracm 1{24}}\a\partial^{\un c} \o_{\un c \un a}\partial_{\un b} 
\o^{\un b \un a} -{\fracm 18}m^2\o^{\un a \un b} \o_{\un a \un 
b}\cr
&\,&  {~~~~~~~~~~} -{\fracm 1{24}}\a h\partial^{\un a} \partial^{\un b} 
h_{\un a \un b} \,+\, {\fracm 1{12}}(\a-1)\partial_{\un b} \o^{\un b 
\un a} \partial^{\un c} h_{\un c \un a} \,-\,{\fracm 14}  m(\partial_{\un a}  
\l_{\un b} ) h^{\un a \un b} \cr
&\,&  {~~~~~~~~~~} -{\fracm 13}\a(\partial\cdot D )(\partial \cdot A)
\,+\, {\fracm 23}D^{\un a} \Box A_{\un a} \,-\,  m^2D\cdot A
\,-\, {\fracm 1{48}}\e_{\un a \un b \un c \un d}\partial^{\un b}\o^{\un c
\un d}\Box A^{\un a}  
\cr
&\,&  {~~~~~~~~~~} + m\h\partial\cdot A
\,-\,{\fracm 12}  m\c\Box \partial\cdot A
\,+\, {\fracm 1{24}}\e_{\un a\un b\un c\un d}\partial^{\un b}\o^{\un c\un d} D^{\un a}   
\,+\,{\fracm 14}  m(\partial_{\un a} \l_{\un b} )\o^{\un a \un b}\cr
&\,&  {~~~~~~~~~~} -{\fracm 23} D\cdot D
\,+\, m\c\partial\cdot D \,+\,{\fracm 18} m(\partial\cdot
\l)h \,-\,{\fracm 13}\a(\partial\cdot F)(\partial\cdot {\Bar 
F})\cr
&\,&  {~~~~~~~~~~} - m^2F\cdot {\Bar F} \,+\,  
m\left(\Bar \j\partial\cdot F \,+\, \j\partial\cdot \Bar F\right)\cr
&\,&  {~~~~~~~~~~} -\a\left( 2 \h\Box \c+2 \j \Box {\Bar \j}
-{\fracm 12}\c\Box^2 \c \,+\, {\fracm 14}\l^{\un a }\Box \l_{
\un a  }\right)\cr
&\,&  {~~~~~~~~~~} +\, (\a-{\fracm 12}) \left( +4 {\Bar \j}\Box\j
\,+\, 4 \h\h \,+\,{\fracm 14} (\partial \cdot \l\right)^2)\cr
&\,&  {~~~~~~~~~~} +{\fracm 32}(1-2\a) m^2 ( 2 \h\c \,+\, 2\j
{\Bar \j} \,-\, {\fracm 12}\c\Box \c
\,+\,{\fracm 14} \l\cdot \l) \Big\} ~~~.
\eea
This action is quite deceiving.  Initially, it appears to possess
higher derivatives acting on the $A^{\un a}   $ field.  Secondly, 
the role of the antisymmetric second rank tensor $\o_{\un a  \un b}$ 
is unclear.  Finally, it would be instructive to show that the equations 
of motion for the fields $h_{\un a \un b}$ and $A_{\un a}$ are
compatible with the subsidiary conditions defining the massive irreducible
spin-2 and spin-1 representations of the Poincar\'e group respectively.

To clarify the situation one can consider the equations of motion for the
action (\ref{compact}). Then one can show that the fields $\c$, $\h$, $\j$, $\Bar
\j$, $F_{\un a}$, $\Bar F_{\un a}$, $D_{\un a}$ and $\partial\cdot A$ all vanish due
to the equations of motion.  Additionally, the $D_{\un a} $-field equation
of  motion yields the following constraint:
\bea
\label{EomDa}
(\Box-{\fracm 32} m^2) \, A^{\un d} ~=~ +{\fracm 1{16}}\e^{\un a 
\un b \un c \un d} \, \partial_{\un a} \o_{\un b \un c} ~~~.
\eea
The equation of motion for $A_{\un a}   $
has the form of the d'Alembertian of equation (\ref{EomDa}).  Thus, the higher
derivative terms are inconsequential, because the same information is
encoded in (\ref{EomDa}).  Up to this point, the equations of motion for $\l_{
\un a}$, $\o_{\un a \un b}$, $h$ and $h_{\un a \un b}$ have not been
used.  The equation of motion of $h$, the divergence of the $\l_{\un a}$
equation of motion, and the second divergence of the $h_{\un a \un b}$
equation of motion imply that $h=\partial^{\un a}\l_{\un a}=\partial^{\un a}  
\partial^{\un b} h_{\un a \un b}=0$.  With this the equation of motion for
$\l_{\un a}$, and the divergence of the equations of motion for
$h_{\un a \un b}$ and $\o_{\un a  \un b}$ imply that $\l_{\un a} =
\partial^{\un a} h_{\un a \un b}=\partial^{\un a} \o_{\un a \un 
b}=0$.

Taking into account the above equations we obtain the equation of motion
for $\o_{\un a \un b}$:
\bea
\label{EomO}
-{\fracm 1{12}}\e_{\un a  \un b \un c \un d}\Box\partial^{\un c} A^{
\un d} \,+\, \left({\fracm 13}\Box-m^2\right)\o_{\un a  \un b} ~=~ 0 ~~~,
\eea
and the mass shell condition for $h_{\un a \un b}$:
\bea
(\Box \,-\, m^2) \, h_{\un a  \un b} ~=~ 0 ~~~.
\eea
Equations (\ref{EomDa}) and (\ref{EomO}) imply both the mass-shell condition for $A^{\un 
a}$, and that $\o_{\un a \un b}$ is the dual of the field strength of
$A^{\un a}   $:
\bea
(\Box \,-\, m^2) \, A^{\un a} ~=~0  ~~~, ~~~
\o_{\un k \un l} \, =\, -{\fracm 18}\e_{\un k \un l \un a \un
b} \, \partial^{\un a}   A^{\un b} ~~~.
\eea
As a result, the field $\o_{\un a  \un b}$ is not independent, it is
expressed in terms of the field $A_{\un a}   $. Thus, although the action (\ref{compact})
looks very complicated and apparently contains higher derivative
terms, the corresponding equations of motion reduce to known equations for
irreducible massive spin-1 and spin-2 representations of the Poincar\'e
group as discussed in the introduction.

\noindent{\bf 6.} 
In summary, in this letter we have presented a new 4D, $N$=1 supersymmetric model describing the propagation of free massive spin-2, spin-3/2 and spin-1 fields.
This model is a supersymmetric
extension of the Fierz-Pauli theory. We have shown that the model is
completely formulated in superfield terms and described by real vector
$V_{\un a}$ and scalar $V$ superfields. The superfield $V_{\un a}$ is
dynamical, while the superfield $V$ is auxiliary.  The role of $V$ is to ensure the
existence of a Lagrangian formulation of the model compatible with the
correct subsidiary conditions defining the irreducible massive
superspin-${\fracm 32}$ representation of the Poincar\'e supergroup. The
corresponding superfield action is given in explicit form (\ref{finact}).

We have actually found two 4D, $N$ = 1 superfield actions that describe 
a massive spin-2 quantum. In the case where $\a$ = 1, the $m \to 0$ and $V\to 0$
limit of our massive action goes smoothly over to become the linearized action 
of old minimal superfield supergravity.  
In the other case where $\a=0$, the $m\to 0$ and $V\to 0$ limit leads to a new off-shell version of linearized superfield supergravity which does not  correspond to either the old minimal, non-minimal or new minimal versions.
This suggests that a higher spin
supersymmetric Higgs-like interpretation of these versions may be possible.  
Another question we must ponder is whether there exist
other massive versions that smoothly connect to other off-shell
versions of superfield supergravity.
It would be 
extremely interesting to construct examples of other theories with various 
superspins, to develop a systematic theory corresponding to arbitrary 
superspin, to couple such models to external supergravity (e.g. to 
construct the massive higher spin supersymmetric models in AdS space)
and investigate the possibility of finding massive higher spin field models
with extended supersymmetry.

\vspace{0.5cm}
${~~~}$ \newline
${~~~~~~~~~}$``{\it {Certitude is not the test of certainty. We have been cocksure of many
things that are not so.}}'' \newline
${~~~~~~~~~~~}$ -- Holmes, Oliver Wendell

${~~~}$

\noindent{\bf Acknowledgements} \newline We are very grateful to N. Berkovits and S.M. Kuzenko for useful comments and discussions. The work of I.L.B. was supported 
in part by the INTAS Grants No 991-590 and No 00-00254. I.L.B. is very grateful to
University of Maryland, where part of this work has been undertaken, for
support of his visit and warm hospitality. Also the work of I.L.B.
is done under the auspices of RFBR project No.\ 02-02-16642.

\end{document}